\documentclass[epj]{svjour}
\usepackage{amsfonts}
\usepackage{amsmath}
\usepackage{amssymb,bbold}
\usepackage{kantlipsum,widetext}
\RequirePackage{graphicx}
\RequirePackage{mathptmx}
\RequirePackage{flushend}
\RequirePackage[colorlinks,citecolor=blue,urlcolor=blue,linkcolor=blue]{hyperref}
\RequirePackage[numbers,sort&compress]{natbib}
\RequirePackage{color}

\journalname{Eur. Phys. J. Plus}
\begin{document}

\title{Scalar bosons with Coulomb potentials in a cosmic string background: Scattering and bound states
}

\titlerunning{Scalar bosons with Coulomb potentials in a cosmic string background}        

\author{Francisco A Cruz Neto\inst{1}\and Franciele M da Silva\inst{2}\and Luis C N Santos\inst{3}\and Luis B. Castro\inst{1,4}
}

\institute{Departamento de F\'{\i}sica, Universidade Federal do Maranh\~{a}o (UFMA), Campus Universit\'{a}rio do Bacanga, 65080-805, S\~{a}o Lu\'{\i}s, MA, Brazil. \and
Departamento de F\'{i}sica - CFM, Universidade Federal de Santa Catarina (UFSC), 88040-900, Florian\'{o}polis, SC, Brazil.\and
Departamento de F\'{\i}sica - CCEN, Universidade Federal da Para\'{\i}ba (UFPB), 58051-970, Jo\~{a}o Pessoa, PB, Brazil. \and
Departamento de F\'{\i}sica e Qu\'{\i}mica, Universidade Estadual Paulista (UNESP), Campus de Guaratin\-gue\-t\'{a}, 12516-410, Guaratinguet\'{a}, SP, Brazil, \email{lrb.castro@ufma.br,luis.castro@pq.cnpq.br}
}

\date{Received: date / Accepted: date}

\abstract{
The relativistic quantum motion of scalar bosons under the influence of a
full vector (minimal $A^{\mu}$ and nonminimal $X^{\mu}$) and scalar ($V_{s}$) interactions
embedded in the background of a cosmic string is explored in the context of
the Klein-Gordon equation. Considering Coulomb interactions, the
effects of this topological defect in equation of motion, phase shift and
S-matrix are analyzed and discussed. Bound-state solutions are obtained from
poles of the S-matrix and it is shown that bound-state solutions are
possible only for a restrict range of coupling constants.
}

\maketitle

\section{Introduction}
\label{intro}

Scalar bosons usually are represented by the Klein-Gordon (KG) equation.
Vector and scalar interactions are considering by replacing (usual
substitutions) 
\begin{eqnarray}
p^{\mu}&\rightarrow & p^{\mu}-A^{\mu}\,,  \label{acomim1} \\
M & \rightarrow & M+V_{s}\,.  \label{acomim2}
\end{eqnarray}
\noindent With this substitutions, a variety of relativistic and
nonrelativistic effects can be studied. The vector interaction $A^{\mu}$
refers to a kind of coupling that behaves like a four-vector under a Lorentz
transformation and it can be associated to electromagnetic interaction. On
the other hand, the scalar interaction $V_{s}$ refers to a kind of coupling
that behaves like a scalar (invariant) under a Lorentz transformation.
Though the scalar interaction finds many of their applications in nuclear
and particle physics, it could also simulate an effective mass in solid
state physics. Due to weak potentials, relativistic effects are considered
to be small in solid state physics, but the relativistic wave equations can
give relativistic corrections to the results obtained from the
nonrelativistic wave equation, therefore the relativistic extension of this
problem is also of interest and remains unexplored.

An additional form of interaction is achieved through the substitution 
\begin{equation}  \label{aconomim}
\vec{p}^{2}\rightarrow \left(\vec{p}+iM\omega r\hat{r}\right)\cdot\left(\vec{%
p}-iM\omega r\hat{r}\right)\,,
\end{equation}
\noindent such interaction is dubbed the KG oscillator. This alternative
form of interaction furnishes a Schr\"{o}dinger equation with the harmonic
oscillator potential in a nonrelativistic scheme \cite{INC106:711:1993} and
spurred great deal of research in last years \cite%
{INC107:1413:1994,CTP42:664:2004,PS77:015003:2008,CTP53:231:2010,IJTP50:228:2011, CTP55:405:2011,PS84:037001:2011,IJTP50:3105:2011,IJMPA27:1250047:2012,JMP55:033502:2014,AP355:48:2015,AP370:128:2016}%
. Considering the KG oscillator as a particular case of other four-vector $%
X^{\mu}$, we can generalize this interaction through the substitution \cite%
{AP378:88:2017} 
\begin{equation}  \label{aconomim2}
p^{\mu}p_{\mu}\rightarrow
\left(p^{\mu}-iX^{\mu}\right)\left(p_{\mu}+iX_{\mu}\right)\,.
\end{equation}
\noindent Note that in contrast to $A^{\mu}$, $X^{\mu}$ is not minimally
coupled and for this reason $X^{\mu}$ is called of nonminimal vector
interaction. Thus, the KG equation with interactions (the most general
Lorentz structure) consists in vectors ($A^{\mu}$ and $X^{\mu}$), and scalar
($V_{s}$) potentials.

Scalar particles in the background space-time generated by a cosmic string
are very interesting systems that have been studied extensively in the
literature in recent years \cite%
{string12,string13,santos2,santos5,Castro1,Castro2}. A natural question
arises, whether the presence of such kind of topological defect can also
influence the behavior of scattering states and bound states of a given
quantum system. In \cite{Castro1}, the quantum dynamics of scalar bosons
embedded in the background of a topological defect were considered. Notably,
in that work it was shown that the energy spectrum associated with the
scalar sector of the Duffin-Kemmer-Petiau (DKP) equation in a cosmic string
space depends on the deficit angle of the conical space-time. The scattering
S-matrix for a fermionic system\ is considered in \cite{string14}, where it
was shown that phase shifts and the normalization factor are influenced by
the presence of the string.

Solutions of relativistic wave equations in curved spaces with these
potentials have been obtained for many systems. For instance, the Cornell
potential \cite{cornell1} is a linear combination of the Coulomb and linear
potentials. It provides a good description of the heavy quark--antiquark
system. In Ref. \cite{cornell2}, relativistic Landau levels in the rotating
cosmic string space-time with the Cornell potential, is considered. It was
shown that the Landau levels of the spinning cosmic string remain the same
even when the internal magnetic flux vanishes. Solutions of the KG equation
with Coulomb-type scalar potential in the cosmic string space-time are
considered in Ref \cite{santos2}. Another example of vector--scalar Coulomb
potentials has been developed in \cite{string12} considering the presence of
a dyon and Aharonov-Bohm magnetic field. In this paper the energy spectra
and the scattering states of the KG equation have been analyzed, it was
shown that the phase shifts depend on the geometry of the space-time.
Recently, solutions of the D-dimensional KG equation via mapping onto the
nonrelativistic one-dimensional Morse potential have been considered \cite%
{AP378:88:2017}. This is a straightforward procedure for finding solutions
of the scalar equation with scalar, vector, and nonminimal coupling
potentials. The influence of a Coulomb-type potential on the KG oscillator
has been investigated by Bakke and Furtado \cite{AP355:48:2015}, the authors
have determined bound-state solutions to the KG equation for both attractive
and repulsive Coulomb-type potentials.

Due to the analogy between disclinations in solids \ and cosmic strings \cite%
{defects}, several results obtained from the analysis of cosmic strings may
be useful in the study of condensed matter systems. This fact is associated to the metric which describes a disclination corresponding to the spatial part of the line element of the cosmic
string. The study of system of fermions in low dimensions play an important role in the context of integer quantum Hall effect in graphene \cite{S315:1379:2007} and topological insulators \cite{RMP82:3045:2010}, among others. It is worthwhile to mention that such problems also have their version for bosonic systems: the integer quantum Hall effect for bosons \cite{PRL110:046801:2013} and symmetry--protected topological (SPT) phase \cite{PRL115:116802:2015}. In this context, bosonic systems has been employed on the study of novel topological semimetals \cite{PRB92:235106:2015}. Other potential applications can be found in Bose--Einstein (BE) condensates and neutral atoms  \cite{defects2,defects3}. Therefore, we believe that the Klein--Gordon equation embedded in a background of cosmic string under external interactions deserves to be more explored. 

The purpose of this paper is to study scattering and bound-state solutions of scalar
bosons with a Coulomb potential with the most general Lorentz structure embedded in a cosmic string background. In this case, the problem is mapped into a Schr\"{o}dinger-like equation with a effective Coulomb-like potential. The Coulomb phase shift and scattering S-matrix are calculated from a Whittaker differential equation via partial wave analysis. The bound-state solutions are obtained from the poles of the S-matrix and the restriction on the potential parameters are discussed in detail. We investigate how the topological and geometric features of the defect affect the energy levels as compared with the flat space-time ($\alpha=1$) and also we are able to reproduce all cases already discussed in the literature as particular cases. Beyond its intrinsic interest, the Coulomb potential (hydrogen atom) in a curved space-time can give us a heuristic look of a plausible mechanism for the detection of gravitational radiation \cite{parker6}. Therefore, ours results shed some light on better understanding of energy level shifts due to topological and geometric features, which in principle enables one to emulate a radiation detector. 

This paper is structured as follows: In Section \ref{sec:cs}, we give a brief review on a cosmic string background. In Section \ref{sec:gkg}, we consider the generalized Klein-Gordon (KG) equation in a cosmic string background. We also consider a generalized Coulomb potential (Section \ref{sec:subsec:gCp}) and we analyze scattering and bound-state solutions (Sections \ref{subsec:ss} and \ref{subsec:bs}, respectively). Additionally, we discuss some particular cases (Section \ref{subsec:pc}). Finally, in Section \ref{sec:conclu} we present our conclusions.

\section{Cosmic string background}
\label{sec:cs}

The cosmic strings are systems that are supposed to be formed during a
symmetry breaking phase transition in the early universe \cite%
{string1,string2,string3,string4,string5,string6,string7}. They are
1-dimensional topological defects and are candidates for the generation of
observable astrophysical phenomena such as gravitational waves and high
energy cosmic rays \cite{string2}. Furthermore, they would also be
associated with galaxy evolution and gravitational lens \cite%
{string2,string3}.These systems have high mass density, of the order of $%
10^{21}$g/cm, and very small thickness, equivalent to the Compton wavelength 
$10^{-29}$cm. Cosmic strings can be straight and have
infinite length or can form a closed loop. Nowadays, there are not a direct
proof of their existence but, we have some indirect evidences that cosmic
string can really exist \cite{string4}. The high-frequency gravitational
waves emitted by cusps and kinks of cosmic strings might be detectable by
the gravitational waves detectors LIGO/VIRGO and LISA \cite{string11}. The
space-time of a cosmic string has cylindrical symmetry such that its line
element is given by \cite{string5} 
\begin{equation}\label{metric_1}
ds^{2}=-A^{2}\left(\rho\right)dt^{2}+d\rho^{2}+C^{2}\left(\rho\right)
d\varphi^{2}+D^{2}\left(\rho\right)dz^{2},\,  
\end{equation}
\noindent where $-\infty <t<+\infty $, $0\leq \rho <+\infty $, $0\leq \varphi \leq
2\pi $ and $-\infty <z<+\infty $. Let us admit that our space-time is
invariant by boosts in the $z$ direction, so that $A^{2}\left(\rho\right)$
must be equal to $D^{2}\left(\rho\right)$. To determine the remaining
functions, $A\left(\rho\right)$ and $C\left(\rho\right) $, we should
make our metric to satisfy the Einstein field equations. By making this, we
will obtain 
\begin{equation}
ds^{2}=-dt^{2}+d\rho ^{2}+\alpha ^{2}\rho ^{2}d\varphi ^{2}+dz^{2}\,.
\label{metric_2}
\end{equation}%
\noindent In this work we will use spherical coordinates, in this way, we consider the
coordinate transformation $\rho =r\sin \theta $ and $z=r\cos \theta $, the
result is the cosmic string space-time, described by the line element 
\begin{equation}
ds^{2}=-dt^{2}+dr^{2}+r^{2}d\theta ^{2}+\alpha ^{2}r^{2}\sin ^{2}\theta
d\varphi ^{2}\,,  \label{metric}
\end{equation}%
\noindent in spherical coordinates $(t,r,\theta ,\varphi )$, where $-\infty
<t<+\infty $, $r\geq 0$, $0\leq \theta \leq \pi /2$ and $0\leq \varphi \leq
2\pi $. The parameter $\alpha $ is associated with the linear mass density $%
\tilde{m}$ of the string by $\alpha =1-4\tilde{m}$ and runs in the interval $%
\left( 0,1\right] $ and corresponds to a deficit angle $\gamma =2\pi
(1-\alpha )$. Note that, in the limit as $\alpha \rightarrow 1$ we obtain
the line element of spherical coordinates.

\section{Generalized Klein-Gordon equation in a cosmic string background}
\label{sec:gkg}

Scalar bosons are represented by the usual Klein-Gordon (KG) equation which
can be generalized to a curved space-time framework. Incorporating the
Klein-Gordon oscillator as a particular case, a generalized
Lorentz-covariant KG equation for a particle of mass $M$ in a curved
space-time under the influence of external vectors ($A^{\mu}$ and $X^{\mu}$),
and scalar ($V_{s}$) fields reads 
\begin{equation}  \label{eqgkg}
\left[ -\frac{1}{\sqrt{-g}}D_{\mu}^{(+)} g^{\mu \nu }\sqrt{-g}D_{\nu}^{(-)}
+\left( M+V_{s}\right) ^{2}\right] \Psi =0,
\end{equation}%
\noindent where 
\begin{equation}  \label{dercov}
D_{\mu}^{(\pm)}=\partial _{\mu }\pm X_{\mu }+iA_{\mu }\,.
\end{equation}
\noindent We can note that in contrast to $A^{\mu}$, the vector potential $%
X^{\mu}$ is not minimally coupled \cite{AP378:88:2017}. Furthermore,
invariance under the time-reversal transformation demands that $A^{\mu}$ and 
$X^{\mu}$ have opposite behaviours. Similarly to the scalar potential, the
nonminimal vector potential does not couple to the charge (invariance under
the charge-conjugation operation), i.e $X^{\mu}$ and $V_{s}$ do not
distinguish particles from antiparticles. At this stage, without loss of
generality we make $\vec{A}=0$ due to the space component of the minimal
coupling can be gauged away for spherically symmetric potentials.

Considering only time-independent and spherically symmetric potentials $%
A_{0}(\vec{r})=V_{v}(r)$, $V_{s}(\vec{r})=V_{s}(r)$, $X_{0}(\vec{r}%
)=V_{0}(r) $ and $\vec{X}(\vec{r})=V_{r}\hat{r}$, it is reasonable to write
the solution as 
\begin{equation}  \label{wf}
\Psi\left(t,r,\theta,\varphi\right)=\frac{u(r)}{r}f\left(\theta\right)%
\mathrm{e}^{-iEt+im\varphi}\,,
\end{equation}
\noindent where $m=0,\pm 1,\pm 2,\ldots$ and $E$ is the energy of the scalar
boson. Substituting (\ref{wf}) into Eq. (\ref{eqgkg}), we obtain the
following equation 
\begin{equation}  \label{eq_ang}
\left[\frac{1}{\sin\theta}\frac{d}{d\theta}\left(\sin\theta\frac{d}{d\theta}%
\right)-\frac{m^{2}}{\alpha ^{2}\sin ^{2}\theta }\right] f\left( \theta
\right)=\lambda _{\alpha}f\left( \theta \right)\,,
\end{equation}
\noindent for the angular part. The solution $f(\theta)$ can be expressed as
a generalized Legendre functions $P^{\mu}_{\nu}(x)$ \cite{CQG19:985:2002}
and thus we get 
\begin{equation}
\lambda_{\alpha}=l_{\alpha}\left(l_{\alpha}+1\right)\,,
\end{equation}
\noindent where $l_{\alpha}=n+|m_{\alpha}|=l+|m|\left(1/\alpha-1\right)$
with $n$ a non-negative integer, $l=n+|m|$ and $m_{\alpha}=m/\alpha$. Note
that $|m_{\alpha}|$ lies in the range $-l_{\alpha}\leq m_{\alpha}\leq
l_{\alpha}$. Here, $l$ and $m$ are the orbital angular momentum and the
magnetic quantum numbers in the flat space, respectively.

For $r\neq 0$ the radial function $u(r)$ obeys the radial equation 
\begin{equation}  \label{eq_rad}
\frac{d^{2}u}{dr^{2}}+\left[ K^{2}-V_{\mathrm{eff}}-\frac{%
l_{\alpha}\left(l_{\alpha}+1\right)}{r^{2}} \right]u=0\,,
\end{equation}
\noindent where the effective energy $K^{2}$ and the effective potential $V_{%
\mathrm{eff}}$ are expressed by 
\begin{eqnarray}
K^{2} &=& E^{2}-M^{2}\,, \\
V_{\mathrm{eff}} &=& V_{s}^{2}-V_{v}^{2}+2\left(MV_{s}+EV_{v}\right)+\frac{dV_{r}}{dr}+2\frac{V_{r}}{r}+V_{r}^{2}-V_{0}^{2}\,.
\end{eqnarray}
\noindent Therefore, the equation (\ref{eq_rad}) describes the quantum
dynamics of scalar bosons under external interactions in a cosmic string
background, whose solution can be found by solving a Schr\"{o}dinger-like
equation. Note that, if the potentials $V_{s}$, $V_{v}$, $V_{r}$ and $V_{0}$
go to zero at $r\rightarrow\infty$ the solution $u(r)$ has the asymptotic
behavior $\mathrm{e}^{\pm iKr}$ and so we can see that scattering states
only occur if $K\in \mathbb{R}$ ($|E|>M$), whereas bound states might occur
only if $K=\pm i|K|$ ($|E|<M$).

\subsection{Generalized Coulomb potential}
\label{sec:subsec:gCp}

Let us consider the potentials in the form 
\begin{equation}
V_{r}=\frac{\beta_{r}}{r},\quad V_{0}=\frac{\beta_{0}}{r},\quad V_{s}=\frac{%
\alpha_{s}}{r},\quad V_{v}=\frac{\alpha_{v}}{r}\,.  \label{potl}
\end{equation}%
\noindent Substituting (\ref{potl}) in (\ref{eq_rad}), we obtain 
\begin{equation}  \label{eqcou1}
\frac{d^{2}u}{dr^{2}}+\left[ K^{2}-\frac{\alpha _{1}}{r}-\frac{\alpha
_{2}+l_{\alpha}(l_{\alpha}+1)}{r^{2}}\right] u=0 \,,
\end{equation}%
\noindent with 
\begin{eqnarray}
\alpha _{1} &=&2\left(\alpha_{s}M+\alpha_{v}E\right)\,,  \label{lambdaa} \\
\alpha _{2}
&=&\beta_{r}(\beta_{r}+1)+\alpha_{s}^{2}-\beta_{0}^{2}-\alpha_{v}^{2}\,.
\label{etaa}
\end{eqnarray}%
\noindent The equation of motion (\ref{eqcou1}) is precisely the
time-inde\-pendent Schr\"{o}dinger equation for a Coulomb-like potential.
This effective potential has well structure when $\alpha_{1}<0$, which
implies that $E<E_{c}=-M\left(\frac{\alpha_{s}}{\alpha_{v}}\right)$. It is
worthwhile to mention that presence of $\alpha_{v}$ or $\alpha_{s}$ (or
both) is necessary for the existence of bound-state solutions. Additionally,
bound states are expected for $|E|<M$. Therefore, we can conclude that
bound-state solutions are possible only for $\alpha_{s}<|\alpha_{v}|$,
corresponding to energies in the interval $-M<E<E_{c}$.

\noindent On the other hand, using the abbreviations 
\begin{equation}  \label{gammal}
\gamma_{l}=\sqrt{\left( l_{\alpha}+\frac{1}{2} \right)^{2}+\alpha_{2}}\,,
\end{equation}
\begin{equation}  \label{eta}
\eta=\frac{ \alpha_{1}}{2K}\,,
\end{equation}
\noindent and the change $z=-2iKr$, the equation (\ref{eqcou1}) becomes 
\begin{equation}  \label{whit}
\frac{d^{2}u}{dz^{2}}+\left( -\frac{1}{4}-\frac{i\eta}{z}+\frac{%
1/4-\gamma_{l}^{2}}{z^{2}} \right)u=0\,.
\end{equation}
\noindent This second-order differential equation is the called Whittaker
equation, which have two linearly independent solutions $M_{-i\eta,%
\gamma_{l}}(z)$ and $W_{-i\eta,\gamma_{l}}$ behaving like $%
z^{1/2+\gamma_{l}} $ and $z^{1/2-\gamma_{l}}$ close to the origin,
respectively. Owing to $u(0)=0$, one has to consider the solution
proportional to 
\begin{equation}  \label{soluregu}
u(z)=A e^{-z/2}z^{1/2+\gamma_{l}}M\left(
1/2+\gamma_{l}+i\eta,1+2\gamma_{l},z \right)\,.
\end{equation}
\noindent where $A$ is a arbitrary constant, 
\begin{equation}  \label{alpha2vinc}
\alpha_{2}>-1/4\,,
\end{equation}
\noindent and $M\left(a,b,z \right)$ is the confluent hypergeometric
function (Kummer's function) \cite{ABRAMOWITZ1965}. From (\ref{alpha2vinc})
one can obtain an upper limit for $\alpha_{v}$, given by 
\begin{equation}  \label{alpha2max}
|\alpha_{v}|<\left(\alpha_{v}\right)_{\mathrm{max}}=\sqrt{\frac{1}{4}%
+\alpha_{s}^{2}+\beta_{r}(\beta_{r}+1)-\beta_{0}^{2}}\,.
\end{equation}

The asymptotic behavior for large $\vert z \vert$ with a purely imaginary $%
z=-i|\bar{z}|$, where $|\bar{z}|=2Kr$ is given by \cite{JAMES1991} 
\begin{equation}\label{asymp}
M\left(a,b,z\right)\simeq \frac{\Gamma(b)}{\Gamma(b-a)}e^{-\frac{i}{2}\pi
a}|\bar{z}|^{-a}+\frac{\Gamma(b)}{\Gamma(a)}e^{-i\left[|\bar{z}|-\frac{\pi}{2}%
\left(b-a\right)\right]}|\bar{z}|^{a-b}\,.
\end{equation}

\subsection{Scattering states}
\label{subsec:ss}

We can show that for $|z|\gg 1$ and $K\in \mathbb{R}$ the asymptotic
behavior dictated by (\ref{asymp}) implies 
\begin{equation}\label{solassy}
u(r)\simeq \sin \left( Kr-\frac{l\pi }{2}+\delta _{l}\right)\,,  
\end{equation}%
\noindent where the relativistic Coulomb phase shift $\delta _{l}=\delta
_{l}\left( \eta \right) $ is given by 
\begin{equation}
\delta _{l}=\frac{\pi }{2}\left( l+1/2-\gamma _{l}\right) +\mathrm{arg}%
\,\Gamma \left( 1/2+\gamma _{l}+i\eta \right) .  \label{pshift}
\end{equation}%
\noindent For scattering states in spherically symmetric scatterers, the
scattering amplitude can be written as partial wave series 
\begin{equation}
f\left( \theta \right) =\sum_{l=0}^{\infty }\left( 2l+1\right)
f_{l}P_{l}\left( \cos \theta \right) \,,  \label{serie1}
\end{equation}%
\noindent where $\theta$ is the angle of scattering, $P_{l}$ is the Legendre
polynomial of order $l$ and the partial scattering amplitude is $%
f_{l}=\left(e^{2i\delta_{l}}-1\right)/\left(2iK\right)$. From this last
expression one can recognize the scattering S-matrix as $S_{l}=e^{2i%
\delta_{l}}$. With the phase shift (\ref{pshift}), up to a logarithmic phase
inherent to the Coulomb field, we find 
\begin{equation}
S_{l}=e^{i\pi \left( l+1/2-\gamma _{l}\right) }\frac{\Gamma \left(
1/2+\gamma _{l}+i\eta \right) }{\Gamma \left( 1/2+\gamma _{l}-i\eta \right) }%
\,.  \label{fl}
\end{equation}%
\noindent Information about the energies of the bound-state solutions can be
obtained from poles of the S-matrix when one considers $K$ imaginary.

\subsection{Bound states}
\label{subsec:bs}

\begin{figure}[]
\includegraphics[scale=0.47]{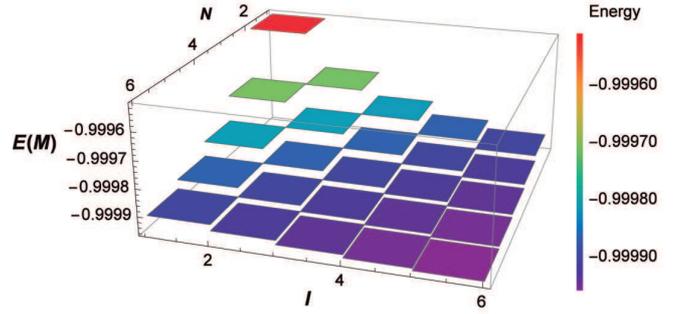}
\caption{Plot of the negative energy spectrum as a function of the quantum
numbers $N$ and $l$ displayed for five different values of $N$ and $l$ with
parameters $\protect\beta _{r}=0.5$, $\protect\beta _{0}=0.6$, $\protect%
\alpha _{s}=0.7,\protect\alpha _{v}=0.8$, $\protect\alpha =0.9 $, and $%
m=1.$}
\label{eminus}
\end{figure}

\begin{figure}[]
\includegraphics[scale=0.6]{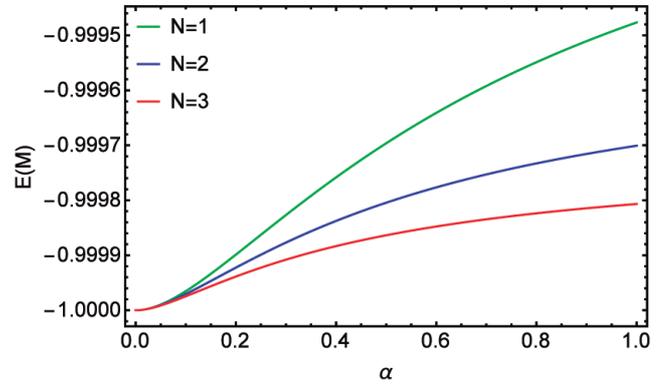}
\caption{Plots of energy versus angular deficit for three values of $N$ with
parameters $\protect\beta _{r}=0.5$, $\protect\beta _{0}=0.6$, $\protect%
\alpha _{s}=0.7,\protect\alpha _{v}=0.8$, $l=1$, and $m=1.$}
\label{ealpha}
\end{figure}

If $K=i|K|$, the S-matrix becomes infinite when $1/2+\gamma _{l}+i\eta =-N$,
where $N=0,1,2,\ldots $, due to the poles of the gamma function in the
numerator of (\ref{fl}), and (\ref{asymp}) implies that $u(r)$ tends to $%
r^{1/2+\gamma _{l}+N}e^{-|K|r}$ for large $r$. Therefore, bound-state
solutions are possible only for $\alpha _{1}<0$ and the spectrum is
expressed as 
\begin{equation}
E=\frac{M}{1+\left( \frac{\alpha _{v}}{\mu }\right) ^{2}}\left[ -\frac{%
\alpha _{s}}{\mu }\frac{\alpha _{v}}{\mu }\pm \Delta \right] \,,
\label{enerq}
\end{equation}%
\noindent where 
\begin{equation}
\Delta =\sqrt{1+\left( \frac{\alpha _{v}}{\mu }\right) ^{2}-\left( \frac{%
\alpha _{s}}{\mu }\right) ^{2}}\,,  \label{delta111}
\end{equation}%
\noindent with $\mu =N+\frac{1}{2}+\gamma _{l}$. Note that the condition $%
\alpha _{s}<|\alpha _{v}|$ guarantees $E$ real. Considering the expression (\ref%
{alpha2max}), we can conclude that bound-state solutions are possible only
if the following constrain on potential parameters is satisfied 
\begin{equation}
\alpha _{s}<|\alpha _{v}|<\sqrt{\frac{1}{4}+\alpha _{s}^{2}+\beta _{r}\left(
\beta _{r}+1\right) -\beta _{0}^{2}}\,.  \label{rangealphav}
\end{equation}%
In this case, $M\left( -N,b,z\right) $ is proportional to the generalized
Laguerre polynomial $L_{N}^{(b-1)}(z)$ \cite{ABRAMOWITZ1965}, and so one
can write the solution as 
\begin{equation}
u(r)\propto r^{\gamma _{l}+1/2}e^{-\frac{|\alpha _{1}|r}{2\mu }%
}L_{N}^{(2\gamma _{l})}\left( \frac{|\alpha _{1}|r}{\mu }\right) \,.
\label{solutionu}
\end{equation}%
Figure \ref{eminus} shows the numerical values of the negative energy
spectrum as a function of the quantum numbers $N$ and $l$. It is noticeable that for a
given set of potential parameters one finds that the lowest quantum numbers
correspond to the highest energy levels and are to be identified with
antiparticle levels. In this case, energy levels for bound-state solutions
can be find in the interval $-M<E<-0.875M$. Also, one sees that the energy
levels tend to $-M$ as quantum numbers increase. The energy levels could
sink into the negative continuum but this couldn't menace the
single-particle interpretation of KG equation since one has antiparticle
levels plunging into the antiparticle continuum. In Fig. \ref{ealpha}, we illustrate the behavior of the energy as a function of $\alpha$ for three different values of $N$. Figure \ref{ealpha} clearly shows the effects of $\alpha$ on the energy levels for fixed values of $N$, $l$ and $m$, one can see that the energy $|E|$ decreases as $\alpha$ increases.  

\subsection{Particular cases}
\label{subsec:pc}

Here, we reproduce some well-known particular cases, already discussed in
the literature. The case of a pure nonminimal vector potential $X^{\mu}$ is not considered because it does not furnish bound-state solutions.

\subsubsection{Minimal vector Coulomb potential}

For $\alpha_{s}=\beta_{r}=\beta_{0}=0$, bound-state solutions are possible
only for $\alpha_1=2\alpha_{v}E<0$, i.e., when $E\gtrless0$ and $%
\alpha_{v}\lessgtr0$, and so the expression (\ref{enerq}) reduces to 
\begin{equation}  \label{energi_v}
E=-\frac{sgn(\alpha_{v})M}{\sqrt{1+\frac{\alpha_{v}^{2}}{(N+1/2+%
\gamma_{l,v})^{2}}}}\,,
\end{equation}
\noindent where $\gamma_{l,v}=\sqrt{\left(l_{\alpha}+\frac{1}{2}%
\right)^{2}-\alpha_{v}^{2}}$. For $\alpha=1$ (flat space-time) this last
result is exactly the expression for the energies for the KG equation with
minimal vector Coulomb potential \cite{GREINER1990}.

\subsubsection{Scalar Coulomb potential}

For $\alpha_{v}=\beta_{r}=\beta_{0}=0$, bound-state solutions are possible
only for $\alpha_1=2\alpha_{s}M<0$, i.e., when $\alpha_{s}<0$ and so the
expression (\ref{enerq}) reduces to 
\begin{equation}  \label{energi_s}
E=\pm M\sqrt{1-\frac{\alpha_{s}^{2}}{(N+1/2+\gamma_{l,s})^{2}}}\,,
\end{equation}
\noindent where $\gamma_{l,s}=\sqrt{\left(l_{\alpha}+\frac{1}{2}%
\right)^{2}+\alpha_{s}^{2}}$. For $\alpha=1$ (flat space-time) this last
result is exactly the expression for the energies for the KG equation with
scalar Coulomb potential \cite{GREINER1990}.

\subsubsection{Mixed scalar-vector Coulomb potential}

For $\beta_{r}=\beta_{0}=0$, bound-state solutions are possible only for $%
\alpha_{1}=2\left(\alpha_{s}M+\alpha_{v}E\right)<0$, i.e., when $%
\alpha_{s}<|\alpha_{v}|$ and so the expression (\ref{enerq}) reduces to 
\begin{equation}  \label{energi_sv}
E =M\frac{ -\frac{\alpha_{s}}{\nu}\frac{\alpha_{v}}{\nu}\pm \sqrt{1+\left(%
\frac{\alpha_{v}}{\nu}\right)^{2}-\left(\frac{\alpha_{s}}{\nu}\right)^{2}} }{%
1+\left(\frac{\alpha_{v}}{\nu}\right)^{2}} \,,
\end{equation}%
\noindent where $\nu=N+\frac{1}{2}+\sqrt{\left(l_{\alpha}+\frac{1}{2}%
\right)^{2}+\alpha_{s}^{2}-\alpha_{v}^{2}}$. For $\alpha=1$ (flat
space-time) this last result is exactly the Eq. (22) of Ref.\cite%
{PRC91:034903:2015}.

\section{Conclusions}
\label{sec:conclu}

We studied the relativistic quantum motion of scalar bosons with a potential with the most general Lorentz structure embedded in the background of a cosmic string via Klein--Gordon equation. Considering Coulomb interactions for the full vector ($A^{\mu}$ and $X^{\mu}$) and scalar ($V_{s}$), this problem was mapped into a Schr\"{o}dinger-like equation with a effective Coulomb-like potential and we showed that the scattering and bound-state solutions can be studied by solving a Whittaker differential equation. 

For scattering solutions ($|E|>M$), the Coulomb phase shift and scattering S-matrix was calculated as a function of the potential parameters and the angular deficit of the cosmic string background $\alpha$. In this case, the minimal vector interaction satisfies the constraint $|\alpha_{v}|<\left(\alpha_{v}\right)_{\mathrm{max}}$ [Eq. (\ref{alpha2max})]. The poles of the S-matrix when $K\rightarrow i|K|$ provided bound-state solutions only for $\alpha_{1}<0$. From this condition, we concluded that the presence of $\alpha_{v}$ or $\alpha_{s}$ (or both) is necessary for the existence of bound-state solutions. The presence of a pure nonminimal vector potential $X^{\mu}$ does not furnish bound-state solutions. Furthermore, we found that bound-state solutions are possible only for $\alpha_{s}<|\alpha_{v}|<\left(\alpha_{v}\right)_{\mathrm{max}}$, corresponding to energies in the interval $-M<E<-M\left(\frac{\alpha_{s}}{\alpha_{v}}\right)$ and also we found that the eigenfunctions are expressed in terms of the generalized Laguerre polynomials. We also showed that the discrete set of energies $|E|$ for this background decreases as $\alpha$ increases. The results obtained in this work are consistent in the limit of flat space-time ($\alpha=1$) with those found in the literature.

Indeed, it is expected that the energy of quantum systems in this space-time carries information about the local features of the background space-time in which the system is placed \cite{hcurvo1}. In this way, it was suggested that the hydrogen atom in a curved space-time can be used as a radiation detector, i.e., the existence of atomic hydrogen in the vicinity of sources of gravitational radiation carry a signature on the energy level shifts and could thus provide a plausible mechanism for detection of gravitational radiation \cite{parker6}. Beyond investigating how the topological and geometric features of the defect affect the energy levels as compared with the flat space-time ($\alpha=1$) our results can be seen as a first step (toy model) for a better understanding of energy level shifts, which enables one to emulate a detector of gravitational waves. This future application is currently under study and will be reported elsewhere.

\begin{acknowledgement}
This work was supported in part by means of funds provided by CNPq, Brazil, Grants No. 307932/2017-6 (PQ) and No. 422755/2018-4 (UNIVERSAL), S\~{a}o Paulo Research Foundation (FAPESP), Grant No. 2018/20577-4 and CAPES, Brazil.
\end{acknowledgement}

\bibliographystyle{spphys}       
\bibliography{cosmic_kg_coulomb}

\end{document}